\documentclass[prb,twocolumn,letterpaper,groupedaddress]{revtex4}
\usepackage{epsfig}
\usepackage{graphicx}
\usepackage{rotating}
\usepackage{multirow}
\usepackage{mathptmx}
\usepackage{amsmath,amssymb,amsfonts}

\begin{document}

\title{Strain-induced isosymmetric phase transition in BiFeO$_3$}
\author{Alison J. Hatt}
\author{Nicola A. Spaldin }
\affiliation{Materials Department, University of California, Santa Barbara, CA
93106-5050}
\author{Claude Ederer}
\affiliation{School of Physics, Trinity College, Dublin 2, Ireland}

\date{\today}

\begin{abstract}
We calculate the effect of epitaxial strain on the structure and properties of
multiferroic bismuth ferrite, BiFeO$_3$, using first-principles density functional
theory.  We investigate epitaxial strain corresponding to an (001)-oriented substrate and find that, while small strain causes only quantitative changes in behavior
from the bulk material, compressive strains of greater than 4\% induce an isosymmetric
phase transition accompanied by a dramatic change in structure. In striking contrast to
the bulk rhombohedral perovskite, the highly strained structure has a $c/a$ ratio of
$\sim$1.3 and five-coordinated Fe atoms. We predict a rotation of polarization from [111]
(bulk) to nearly [001], accompanied by an increase in magnitude of $\sim$50\%, and a
suppression of the magnetic ordering temperature.  Our calculations indicate critical
strain values at which the two phases might be expected to coexist and shed light on
recent experimental observation of a morphotropic phase boundary in strained BiFeO$_3$.
\end{abstract}

\maketitle

\section{Introduction}

BiFeO$_3$ (BFO) has been widely studied for its room temperature
multiferroic properties, in which the electric polarization is coupled
to antiferromagnetic order, allowing for manipulation of magnetism by
applied electric fields and vice versa
\cite{Neaton_et_al:2005,Ederer/Spaldin_PRB:2005,Wang_et_al:2003,
  Zhao_et_al:2006,Lebeugle_et_al:2008}. In its bulk form, BFO occurs
in the $R3c$ space group\cite{Kubel/Schmid:1990}, but it is often studied in the form of thin
films where it is subject to an epitaxial constraint. Such constraints
impose coherency and strain that in general distort the bulk structure
and/or stabilize phases not present in the bulk
material\cite{Biswas_et_al:2001,Schlom_et_al:2007,Balasubramaniam_et_al:2007}.
Indeed, several groups have reported stabilization of a nearly tetragonal
phase in highly strained BFO films grown on LaAlO$_3$ substrates with a
giant $c/a$ ratio close to 1.3 \cite{Bea_et_al:2009, Zeches_et_al:2009}.  A similar result was found by earlier theoretical
studies that constrained BFO to tetragonal $P4mm$ symmetry, thus preventing
the rotational distortions found in the $R3c$ bulk
phase\cite{Ederer/Spaldin_PRL:2005,Ricinschi/Yun/Okuyama:2006}.  Within
this constraint the theoretical ground state of BFO has a massively reduced in-plane
lattice parameter of 3.67\AA\ and a ``super-tetragonal'' unit cell with
$c/a$=1.27. This $P4mm$-constrained structure has an enhanced polarization
roughly 1.5 times that of the bulk single crystal, a remarkable increase
that has also been reported experimentally, though not widely reproduced
\cite{Yun_et_al:2006}.  Lisenkov {\em et al.} recently used model
calculations to suggest that the $P4mm$ phase can also be stabilized by
electric fields, and they identified an additional phase intermediate to
the strained-bulk phase and $P4mm$ induced by application of a
[00$\bar{1}$] electric field \cite{Lisenkov/Rahmedov/Bellaiche:2009}. The
most recent experimental results, from B\'{e}a {\em et al.} and Zeches {\em
et al.}, have resoundingly confirmed the existence of a second phase with
giant $c/a$ in BFO thin films grown on (001) LaAlO$_3$ but suggest a
monoclinically distorted space group $Cc$
\cite{Bea_et_al:2009,Zeches_et_al:2009}.   In contrast to earlier
experimental studies, B\'{e}a {\em et al.} only found a modest enhancement
of the polarization.  Intriguingly, Zeches {\em et al.} observed the
coexistence of the high strain ``super-tetragonal'' phase and a low strain
bulk-like phase with defect- and dislocation-free interfaces between phase
domains, in spite of a large difference in out-of-plane lattice parameters.

In this work we use ab initio calculations to investigate the effect
of epitaxial strain on BFO without the additional imposed symmetry
constraints used in earlier first-principles studies. We address the
most widely used (001)-oriented cubic substrates and cover a range of
strains encompassing all experimentally accessible states. Our
calculations reveal a strain-induced phase transition at $\sim 4.5$\%
compressive strain, consistent with recent experimental reports
\cite{Bea_et_al:2009,Zeches_et_al:2009} and in addition show that the
transition is {\it isosymmetric}. We reported these basic results in Ref.
\onlinecite{Zeches_et_al:2009}, and in the present work we provide the complete theoretical background and analysis. Our analysis of the isosymmetric
behavior in the transition region suggests an explanation for several experimental results
reported in Ref.~\onlinecite{Zeches_et_al:2009}; namely, the coexistence of bulk-like and
super-tetragonal phases in films grown on LaAlO$_3$ substrates,
 and the reversible movement of domain walls in these same films by electric field. In addition, our calculations of the magnetic properties of the super-tetragonal phase point to
magnetic behavior that is distinct from the rhombohedral phase.
Finally, in the low strain regime, we confirm that
epitaxial strain has only a small quantitative effect on the
properties of rhombohedral BFO \cite{Ederer/Spaldin_PRL:2005}.

\begin{figure}[hb]
\epsfig{file=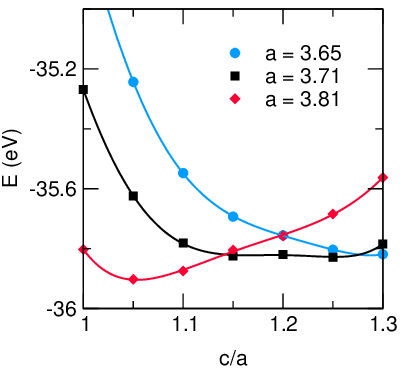,width=4.25cm}
\epsfig{file=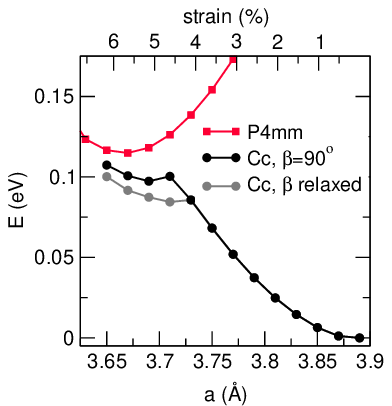,width=4.25cm}
\caption{{\em Left:} Total energy per formula unit as a function of
  $c/a$ ratio for three constrained $a$ parameters. {\em Right:}
  Energy per formula unit relative to bulk BFO for the ground state
  structures as a function of $a$ for $P4mm$, un-relaxed
  monoclinic angle ($\beta$=90$^\circ$), and relaxed
  monoclinic angle.}
\label{E-c}
\end{figure}

\section{Computational Method}

We perform density functional calculations using the local spin
density approximation plus Hubbard $U$ (LSDA+$U$) approach as
implemented in the software package VASP
\cite{Kresse/Furthmuller:1996}. We use an effective $U$ of 2~eV, which
has been shown to give a good description of bulk properties of BFO
\cite{Neaton_et_al:2005}. We use the projector augmented wave method
\cite{Kresse/Joubert:1999}, the default VASP potentials (Bi\_d,
Fe\_pv, O), a 5$\times$5$\times$5 Monkhorst-Pack k-point mesh, and a
500~eV energy cutoff. Spin-orbit coupling is not included in these
calculations and unless otherwise noted we impose the bulk G-type
antiferromagnetic order. Electric polarization is calculated using the
Berry phase method \cite{King-Smith/Vanderbilt:1993,
  Vanderbilt/King-Smith:1993}.

To address the effect of epitaxial strain, we use a 10 atom unit cell with lattice
vectors $\vec{a}_1 = (a, a, 0)$, $\vec{a}_2 = (\Delta, a+\Delta, c)$, and $\vec{a}_3 =
(a+\Delta, \Delta, c)$. This unit cell can accommodate the alternating rotations of the
FeO$_6$ octahedra found in the bulk $R3c$ structure, while enforcing the formation of a
square lattice within the $x$-$y$ plane, corresponding to the epitaxial constraint
imposed by a (001) oriented cubic substrate with in-plane lattice constant $a$.
Furthermore, it allows us to relax the out-of-plane lattice parameter $c$ and a possible
monoclinic tilt $\beta$ of the pseudo-cubic perovskite unit cell ($\tan\beta =
c/\Delta$). For $a=c=3.89$~\AA\ and $\Delta=0$ one obtains the lattice vectors of the
relaxed $R3c$ structure of bulk BFO, albeit with the rhombohedral angle fixed to
60$^\circ$. Note that since the relaxed value of this angle for $U_{\rm eff}=2$~eV is
59.99$^\circ$ (see Ref.~\onlinecite{Neaton_et_al:2005}) this is barely a constraint.  For
computational simplicity we first consider the case $\Delta=0$, i.e. no monoclinic
distortion of the perovskite unit cell ($\beta=90^\circ$); this constraint is relaxed
later. 

\section{Results and Discussion}

\subsection{Energetics and Structure}

We start by examining how the energy varies with out-of-plane lattice parameter for strained BFO.  For in-plane lattice
parameters corresponding to compressive strains up to 6.2\% we vary $c/a$ and, for each
$c/a$ value, we relax all internal coordinates until the Hellman-Feynman forces are no
larger than 1 meV/\AA\ on any atom. The internal coordinates are initialized according to
the relaxed bulk $R3c$ structure, resulting in space group symmetry $Cc$.
Fig.~\ref{E-c}(left) shows the resulting total energy as a function of $c/a$ for three
representative cases.  For $a=3.81$~\AA, corresponding to a moderate compressive strain
of 2.06~\% relative to the LSDA bulk value (3.89~\AA), a single energy minimum is
observed at $c/a \sim 1.05$, i.e. relatively close to unity. For $a=3.65$~\AA, i.e.
6.17~\% compressive strain, we again find a single energy minimum, but this time at a
very large $c/a$ of almost 1.3. In contrast, for an intermediate value of $a=3.71$~\AA\
(4.63~\% compressive strain), the energy curve is almost flat between $c/a \sim 1.1$ and
1.3. Our calculations indicate the presence of two minima at this intermediate $a$, a global minimum at large
$c/a \approx 1.25$ and a local minimum at smaller $c/a \approx 1.15$. Further
calculations would be required to fully resolve the energy curve in this region.  

\begin{figure}[t]
\epsfig{file=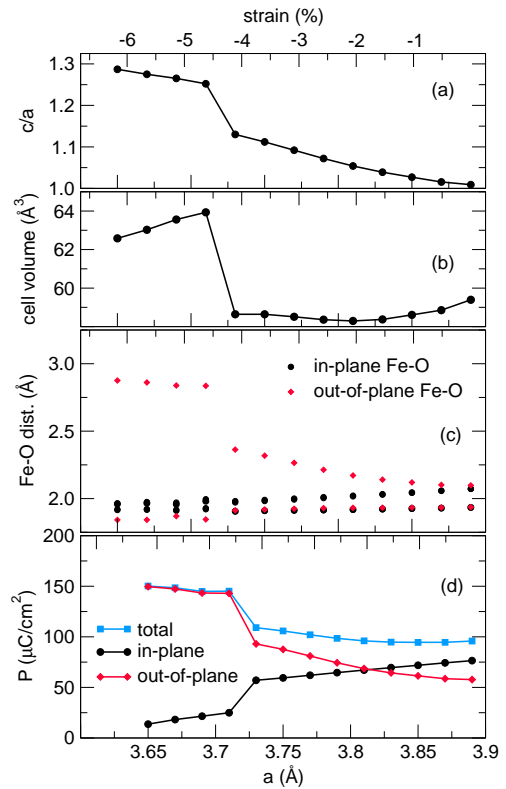,width=6.5cm}
\caption{Structural parameters as functions of lattice parameter/strain relative to the bulk LSDA+$U$ lattice parameter. (a) $c/a$ of the pseudocubic cell; (b) volume per formula unit; (c) Fe-O bond lengths for in-plane and out-of-plane bonds; (d) total polarization and components lying in-plane and out-of-plane.  All values are for constrained monoclinic angle, $\beta$=90$^\circ$.}
\label{order-params}
\end{figure}

We then use polynomial fits to extract the value of $c/a$ that minimizes the total energy
for fixed $a$, and relax all atoms again at the obtained $c/a$ ratio, still maintaining
$\beta = 90^\circ$.  The total energy of the resulting structures is shown in
Fig.~\ref{E-c} (right), relative to the energy of bulk $R3c$.  The corresponding
evolution of various structural parameters ($c/a$ ratio, unit cell volume, Fe-O
distances) and of the electric polarization as a function of strain is shown in
Fig.~\ref{order-params}. All of these quantities exhibit a sharp discontinuity between
$a=3.71$\AA\ and $a=3.73$\AA, i.e.  4-4.5\% strain, where also the total energy depicted
in Fig.~\ref{E-c} (right) has a kink, indicative of a first-order phase transition.

Finally, we perform additional total energy calculations for $\beta \neq 90^\circ$,
determine the optimal $\beta$ for each value of $a$ from a polynomial fit, and again
relax atom positions within this new unit cell.  For $a<3.72$~\AA\ we find a monoclinic
distortion of the unit cell of $\sim$1$^\circ$ to 2$^\circ$ that reduces the energy by
~10 meV/f.u.  compared to $\beta=90^\circ$, as shown in Fig.~\ref{E-c}. The resulting
atomic configuration represents only a minor structural change from the
$\beta$=90$^\circ$ case and we thus neglect this distortion in all further calculations,
only reporting results for $\beta$=90$^\circ$.

\begin{figure}[t]
\epsfig{file=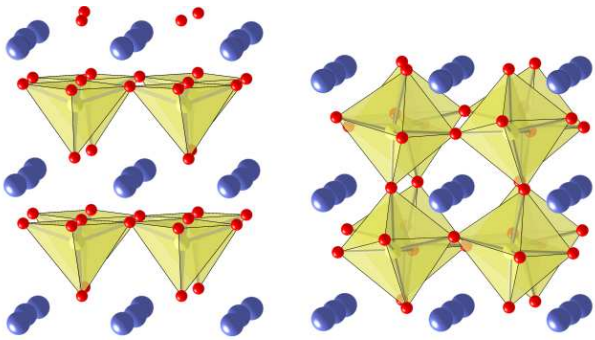,width=7.8cm}
\caption{Isosymmetric phases of $Cc$ BFO: {\em Left:} tetragonal-like
  $T$-phase; {\em Right:} rhombohedral-like $R$-phase.}
\label{unitcell}
\end{figure}

We start our discussion by first characterizing the structures for two extreme cases
representative of the small and large strain regimes, $a=3.85$~\AA\ and $a=3.65$~\AA,
$\sim$1\% and 6\% compressive strain, respectively.  At 1\% strain the relaxed
structure closely resembles the rhombohedral $R3c$ bulk phase, but the epitaxial
constraint causes a monoclinic distortion that lowers the symmetry from $R3c$ to $Cc$.
The corresponding structure is depicted in Fig.~\ref{unitcell} (right). Following the
notation introduced in Ref.~\onlinecite{Zeches_et_al:2009}, we call this the `$R$' phase
to emphasize its similarity to the rhombohedral parent phase. The internal structure
remains similar to the unstrained bulk, with octahedrally coordinated Fe, a ferroelectric
distortion consisting of ionic displacements along the pseudocubic (PS) [111] axis, and
antiferrodistortive rotations of the FeO$_6$ octahedra around [111]$_\text{PS}$.
The [001]$_\text{PS}$ component of the antiferrodistortive rotation increases and the [110]$_\text{PS}$
component decreases as we shrink the in-plane lattice constant, but no qualitative change
occurs for compressive strains up to 4\%. The electric polarization, constrained by
symmetry to lie within the monoclinic glide plane, is almost entirely along
[111]$_\text{PS}$, rotating slightly towards [001]$_\text{PS}$ as compressive strain is
increased (see Fig.~\ref{order-params}d).  Our results are consistent with the growing
body of literature on BFO under moderate compressive
strain\cite{Kim_et_al:2008,Zeches_et_al:2009,Daumont_et_al:2009}

At 6\% strain, the ground state resembles the $P4mm$ structure described previously for
tetragonally-constrained BFO \cite{Ederer/Spaldin_PRB2:2005}.  Using the notation from
Zeches {\em et al.} \cite{Zeches_et_al:2009}, we refer to the high-strain structure as
the `$T$' phase for its similarity to tetragonal $P4mm$.  The Fe undergoes large
displacement towards one of the apical oxygens (see Fig.~\ref{order-params}c), resulting
in a ``super-tetragonal'' structure with five-coordinated Fe, similar to the
coordination of the transition metals in perovskites PbVO$_3$ and BiCoO$_3$
\cite{Belik_et_al:2005, Belik_et_al_CM:2006} (see Fig.~\ref{unitcell}). $P4mm$ symmetry
is broken by antiferrodistortive tilting of the square-pyramidal oxygen cages
(analogous to octahedral tilt) around [110]$_\text{PS}$ by 5.1$^\circ$.
Interestingly, the $T$ phase does not exhibit the [001]$_\text{PS}$ rotations
that have been associated with compressive strain in other perovskites
\cite{Zayak_et_al:2006, Hatt/Spaldin:2009}. This is likely associated with the change in
Fe coordination, which encourages displacement along [001]$_\text{PS}$ 
in response to decreased lattice parameter rather than rotation of the oxygen cages.

As can be seen in Fig.~\ref{E-c} (right), the $T$ phase is significantly lower in energy
than $P4mm$.  Furthermore, it follows from Fig.~\ref{E-c} that at about 5\% strain the
energy reduction obtained by allowing the tiltings  (going from $P4mm$ to $Cc$ at
$\beta=90^\circ$) is even larger than the subsequent reduction from relaxing $\beta$.
These results indicate that the formation of a tetragonal $P4mm$ phase in BFO films
grown on SrTiO$_3$ substrates, as reported e.g. in Ref.  \onlinecite{Yun_et_al:2006}, is
highly unlikely.  

The polarization vector is within the glide plane, as in the low strain regime, but is now strongly rotated to be almost entirely out-of-plane,
along [001]$_\text{PS}$ (Fig.~\ref{order-params}d). In addition, the magnitude (relative
to a $Pm\bar{3}m$ reference structure) increases
from 96 $\mu$C/cm$^{2}$ in the unstrained bulk to 150 $\mu$C/cm$^{2}$.   This is in
contrast to the modest enhancement reported in Ref.~\onlinecite{Bea_et_al:2009}, where
a remnant polarization of 75 $\mu$C/cm$^2$ is reported for Mn-doped films on LaAlO$_3$ 
substrates, compared to 60 $\mu$C/cm$^{2}$ in bulk \cite{Lebeugle_et_al:2007}.  The
discrepancy in enhancement might be a result of the 5\% Mn doping used to reduce
current leakage but it could also indicate incomplete switching of the polarization.
This is reasonable, as we expect the large distortion accompanying the polarization to
be associated with a large coercivity.  Reorientation of the out-of-plane polarization
would require a significant change in bonding for the 5-coordinated structure and it is
possible that the nature of switching is quite different in the super-tetragonal phase
than in the bulk. 

We observe that the total polarization $\vec{P}$ is nearly constant within the low
strain regime (1\%--2\%) but that the out-of-plane projection grows at the expense
of the in-plane projection (Fig.~\ref{order-params}d).  This is in part associated with
the growing $c/a$, but when we correct for that we find that $\vec{P}$ is rotated toward
the [001] direction by several degrees at -1\% strain.  This rotation in the low strain
regime is qualitatively similar to that reported in Ref.~\onlinecite{Jang_et_al:2008},
although we find a much smaller enhancement of the total polarization.  The rotation
across the transition from $R$ to $T$ is much larger:  at $a=$ 3.73~\AA\ the angle
between $\vec{P}$ and [001] is 31.52$^{\circ}$; at $a=$ 3.71~\AA\ it is 9.90$^{\circ}$.
We suspect that this dramatic rotation is responsible for the recently demonstrated
ability to shift the phase boundary with an electric field in BFO films with mixed $T$
and $R$ phases\cite{Zeches_et_al:2009}.  An electric field applied along [001] will
favor an out-of-plane polarization, thus growing the $T$ phase at the expense of $R$.
The properties that are responsible for the side-by-side coexistence of $T$ and $R$
domains, and for the reversible nature of this domain shifting, are addressed later.

\subsection{Octahedral Modes}
 
\begin{figure}[h]
\epsfig{file=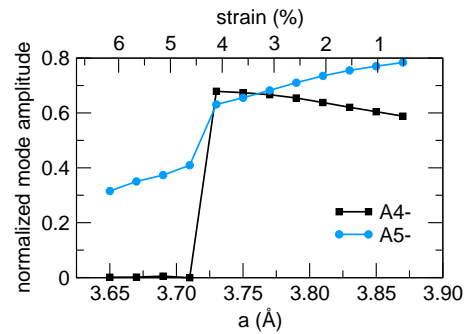,width=6.0cm}
\caption{Displacement mode amplitudes for dominant antiferrodistortive modes.
  A$_{4-}$ and A$_{5-}$ correspond to octahedral rotation and tilt,
  respectively.  }
\label{rot_modes}
\end{figure}

To further quantify the phase transition, we decompose the structural distortion relative
to a $P4/mmm$ reference structure into symmetry irreducible modes using the \sc
ISODISPLACE \rm software \cite{Campbell_et_al:2006}. We find the largest changes as a
function of strain in the antiferrodistortive $A_{4-}$ and $A_{5-}$ modes (see
Fig.~\ref{rot_modes}). These modes correspond to alternating rotations of the FeO$_6$
octahedra around the [001]$_\text{PS}$ and [110]$_\text{PS}$ axes, respectively, and are
therefore a measure of the octahedral rotations and tilts, similar to the analysis
presented in previous work \cite{Zayak_et_al:2006, Hatt/Spaldin:2009}. In fact, the
dependence of rotation and tilt angles (using the terms as defined in
Ref.~\onlinecite{Zayak_et_al:2006}) on lattice parameter are qualitatively identical to
those of the corresponding symmetry adapted modes. As can be seen from
Fig.~\ref{rot_modes}, the transition between $T$ and $R$ phases can be characterized by
the evolution of the A$_{4-}$ and A$_{5-}$ modes as a function of lattice parameter,
which again exhibit pronounced discontinuities around $a=3.71$~\AA. In particular, the
A$_{4-}$ mode vanishes in the $T$ phase. This suggests that the isosymmetric phase
transition is related to the stability of the corresponding phonon mode, which is
unstable for larger in-plane lattice parameters whereas it becomes stable for large
compressive in-plane strain.  The phase transition thus seems to be of similar
microscopic origin as conventional soft mode transitions, albeit with no resulting change
in space group symmetry.

\subsection{Magnetic Coupling}

Finally, we calculate the relative energies of likely magnetic orderings in the $T$
phase. We double the unit cell to allow A- and C-type orderings and re-relax the atom
positions for each fixed magnetic order. We do not optimize the volume for the different
configurations but maintain that of the G-type unit cell. For the $R$ phase, G-type
antiferromagnetic ordering is the most stable magnetic configuration across the whole
strain range, consistent with bulk. In contrast,
we find that in the $T$ phase, G-type and C-type are nearly degenerate, with C-type being
slightly lower in energy; for example, at 6\% strain, the difference is 6 meV per formula
unit \cite{mag_footnote}.  In both G- and C-type, neighboring Fe moments within
(001)$_\text{PS}$ are antiferromagnetically aligned, but whereas for G-type neighboring
moments are also antialigned in the out-of-plane direction, they are ferromagnetically
aligned for C-type.  A-type ordering, with parallel orientation of all magnetic moments
within the same (001)$_\text{PS}$ plane but alternating order in the perpendicular
direction, is strongly unfavorable.

To further quantify this, we map the calculated total energies onto a nearest neighbor
Heisenberg model where $E=-\frac{1}{2}\Sigma_{ij} J_{ij} S_i S_j$, and calculate the
magnetic coupling constants $J_{ij}$ for $S_i=\pm\frac{5}{2}$. We distinguish between
in-plane coupling, $J_\text{in}$, and out-of-plane coupling $J_\text{out}$. For the
$T$-phase at 6\% strain, we find $J_\text{in}=-10$~meV and $J_\text{out}=0.48$~meV,
whereas for the $R$-phase at 1\% strain, $J_\text{in}=-9.8$~meV and
$J_\text{out}=-7.6$~meV. This shows that the increased distance between Fe atoms along
[001]$_\text{PS}$ in the high strain regime strongly reduces the magnetic coupling
strength in that direction, leading to the very similar energies of C- and G-type
magnetic order. The weak magnetic coupling between adjacent (001)$_\text{PS}$ planes is
likely to significantly reduce the magnetic ordering temperature, in spite of the strong
coupling within individual (001)$_\text{PS}$ planes.

\section{Summary and discussion}

In summary, our calculations of the effect of strain on BFO reveal two distinct
structures in the high and low strain regime with a discontinuous first-order transition
between them at $\sim$4.5~\% compressive strain. We note that, as a result of the
constraints imposed by coherence and epitaxy, both phases have the same space group
symmetry; such phase transitions are known are \emph{isosymmetric} and are necessarily
first order\cite{Christy:1995}. Isosymmetric transitions have also been demonstrated in
pressure-induced transitions, as in Fe$_{0.47}$NbS$_2$ and
$\alpha$-PbF$_2$,\cite{Ehm_et_al:2007, Haines/Leger/Schulte:1998} or by temperature
changes, as in $\beta$-YbV$_4$O$_8$ and the fulleride
Sm$_{2.75}$C$_{60}$,\cite{Arvanitidis_et_al:2003, Friese_et_al:2007}.  Large volume
changes appear to be characteristic. For example the pressure-induced spin-state
transition in cerium is accompanied by a 16.5\% volume change.  Indeed BFO also undergoes
a ~9\% change in volume from $R$- to $T$-phase (see Fig.~\ref{order-params}b).  To our
knowledge this is the first example in the literature of a strain-induced isosymmetric
phase transition.  The recent work by Lisenkov {\it et al.}
\cite{Lisenkov/Rahmedov/Bellaiche:2009} showed that isosymmetric phase transitions in
BiFeO$_3$ films may also be induced by electric fields.

While such isosymmetry might enable the coexistence of the two phases recently reported
in strained thin films \cite{Zeches_et_al:2009}, we suggest that the low barrier between
$R$ and $T$ phase, as shown in Fig.~\ref{E-c}, is likely also a requirement.  An $R$
phase film with large compressive strain near the transition region can lower its energy
significantly by relaxing to a slightly larger lattice parameter.  This lattice expansion
would normally create massive dislocations within the crystal, but because of the
unusually flat energy surface near the transition region, it can be accommodated by the
simultaneous development of $T$ domains with decreased lattice parameter, incurring a
relatively small energy penalty.  This model for coexistence in BFO thin films can provide
guidelines for identifying other systems with coexisting phases, namely the presence of
an isosymmetric transition with a low energy barrier between phases.  We also suggest
that the isosymmetric nature of the transition facilitates the reversible movement of
the boundary between $R$ and $T$ phases with an electric field\cite{Zeches_et_al:2009},
as it obviates the need to change symmetry at the transition.  The practical and
technological implications of phase coexistence and morphotropic phase boundary-like
behavior are being actively explored and we hope that the criteria suggested by our
calculations will provide direction for future studies.

\begin{acknowledgments}

We gratefully acknowledge support from the following:  NSF Award Nos. DMR-0820404 and
NIRT-0609377 (NAS and AJH) and Science Foundation Ireland through Contract No.
SFI-07/YI2/I1051 (CE).  Travel support was provided by the International Center for
Materials Research through the IMI Program of the NSF under Award No.  DMR04-09848
(AJH).  Computational resources used include the SGI Altix [Cobalt] system and the
TeraGrid Linux Cluster [Mercury] at the National Center for Supercomputing Applications
under Grant No.  DMR-0940420; CNSI Computer Facilities at UC Santa Barbara under NSF
Grant No. CHE-0321368.; and facilities provided by the Trinity Centre for High
Performance Computing.  We thank R.J. Zeches, M.D. Rossell, L.W Martin, and R.Ramesh for
helpful discussions.  
\end{acknowledgments}

\bibliographystyle{apsrev}

\end{document}